\documentclass[reqno]{amsart}
\usepackage[margin=2cm]{geometry} 
\usepackage{color, graphicx}
\DeclareMathOperator{\pf}{pf}
\DeclareMathOperator{\n}{n}

\newcommand{\DpfDt}{\frac{\Delta \pf}{\Delta T}}
\numberwithin{equation}{section}
\numberwithin{table}{section}
\title{A Model for Tax Evasion with some Realistic Properties}

\date{\today}
\author{Richard Vale}
\email{rval012@aucklanduni.ac.nz}

\begin{document}
\begin{abstract}
We present a discrete-time dynamic model of income tax evasion. The model is solved exactly in the case of a single taxpayer and shown to have some realistic properties, including avoiding the Yitzhaki paradox. The extension to an agent-based model with a network of taxpayers is also investigated.
\end{abstract}
\maketitle
\begin{flushleft}

\section{Introduction}\label{introduction}
\subsection{}
There are various reasons for building a mathematical model of tax evasion. It is interesting to create models which exhibit some features of the real world, as this might help us to understand the mechanisms underlying taxpayer behaviour. It is also useful to have models available for testing the possible effects of new compliance strategies such as changes in tax rates.

\subsection{}
We present a model of income tax which combines new ideas with ideas from existing models. The model is dynamic (evolves in time) and quite simple, having four parameters in the one-taxpayer case. These parameters are the tax rate, the probability of audit, the penalty rate and the savings rate (see Section \ref{model} for the details.) The new model is interesting because it has two realistic features. 

\subsection{}
Firstly, many existing models predict that there should be widespread non-compliance, but in fact many people pay tax. Some authors explain this by introducing a notion of tax morality or proposing that there is a ``social reward" for paying tax which should be included in calculations. Since taxes pay for infrastructure, it makes sense that taxpayers would be motivated by social considerations. Nevertheless, the present paper aims to show that widespread compliance can arise in a population of purely selfish taxpayers.

\subsection{}
Secondly, another strange aspect of tax evasion models is that many of them predict that increasing the tax rate should increase compliance. This does not seem to agree with real life; when taxes are very high, they are viewed as unfair, and people are more reluctant to pay them. In the extreme case, everybody would be expected to evade a tax over 100\%\footnote{We consider only simple models in which a taxpayer's entire income is taxed at a fixed rate. It is not clear that taxpayers with a \emph{marginal} tax rate of 100\% would be expected to evade.} as this is equivalent to confiscation of all income. The problem of models predicting that higher taxes should result in higher compliance is sometimes called the Yitzhaki Problem or Yitzhaki Paradox.

\subsection{}
The model presented in this paper predicts that higher taxes should result in lower compliance, provided that the tax rate exceeds a certain critical value, and a 100\% tax will result in total non-compliance. Again, the emotions of the taxpayers do not need to be taken into account. The relationship between tax rate and non-compliance is derived from economic considerations and taxpayers seek only to maximise their profits.

\subsection{}
In the simplest terms, the reason why the new model predicts widespread compliance is because the tax authority is assumed to collect all unpaid taxes rather than just taxes from the most recent period. Some authors have previously considered models like this, although Seibold and Pickhardt (who call this feature \emph{back-auditing} \cite{Seibold}) comment that it has been `largely neglected in the literature'. The reason why the new model predicts higher non-compliance at higher tax rates is that people will not have enough money to live on if the tax rate is too high.

\subsection{References}
The most famous model of tax evasion is the 1972 Allingham and Sandmo model \cite{AS}. The Yitzhaki Paradox originates in Yitzhaki's 1974 paper \cite{Yitzhaki}. The question of the effect of tax rates on tax compliance in the real world has been much studied.
and a survey  of models, empirical studies and experimental results is \cite{SP}. 
The model introduced in the present paper is related to a model of Klepper, Nagin and Spurr \cite{KNS} in which taxpayers make a decision whether to evade based on their savings rate. However, the model of \cite{KNS} only has two periods ($t=0, 1$.) The original aim of the current paper was to develop a model for a single taxpayer which can be used as an agent in an agent-based model of tax evasion on a network. This was inspired by the agent-based model of Korobow, Axtell and Johnson \cite{KAJ}. The model using a toroidal grid in Section \ref{torus} is closely based on the Ising model of statistical mechanics. The idea of using the Ising model to study tax evasion is due to Zaklan, Westerhoff and Stauffer \cite{Zaklan}. 

\section{A Model of a Single Taxpayer}\label{model}
\subsection{}
In this section we present a model for a single taxpayer. The model turns out to be simple enough to have an explicit solution. First we describe the model in words. There are four parameters: the tax rate $\tau$, the savings rate $k$, the probability of audit $p$ and the penalty rate $\lambda$. We assume $0 < \tau, k, p < 1$ and $\lambda > 1$.

\subsection{Verbal Description}
The taxpayer has an income of $\$1$ and can choose whether to pay the tax of \$$\tau$. If the taxpayer pays the tax, the taxpayer is said to \emph{comply} and receives \$$(1-\tau)$. Otherwise, the taxpayer \emph{evades} and receives $\$1$. The taxpayer saves a proportion $k$ of this income and spends the rest. The taxpayer may then be audited. If audited, the taxpayer must pay back all the tax evaded so far, together with a penalty which is proportional to the amount of tax evaded. Because the taxpayer spends some of his or her income, it may not be possible to recover all of the evaded tax. In this case, the taxpayer pays as much as possible, and then continues with $\$0$ at the next time step.

\subsection{}
The decision whether to evade is based on the profit made from evasion so far. If the profit made from evasion between time $0$ and time $t$ is positive then the taxpayer evades at time $t$, otherwise the taxpayer complies at time $t$.

\subsection{}
The model is a very simplified version of how a real-world tax authority operates. The tax authority, upon uncovering evidence of evasion, attempts to recover all the evaded tax, not just the tax from the most recent period. But this is not always possible, because the taxpayer might have spent some of the evaded tax. Real-world tax agencies might seize the taxpayer's assets, but might still fall short if the evaded tax had been spent on perishable items such as meals, holidays or gambling. Interest is ignored in the model. In real life, the taxpayer might earn interest on savings, but this is balanced by the interest that must be paid on evaded tax. Of course, this is one of many assumptions that simplify the model. For example, we ignore utility and assume that tax is paid on the whole income at the same fixed rate.

\subsection{Detailed Description}
We now give a more detailed mathematical description of the model. Time moves in discrete steps $t = 0,1,2, \ldots$. At time $t$, the taxpayer may or may not evade (the condition for evasion will be given below.) If the taxpayer evades, the income received is $1$. If the taxpayer does not evade, the income received is $1-\tau$. A proportion $k$ of the income is retained. Let $f(t)$ denote the taxpayer's fortune at time $t$. Then
$$
f(t+1) = \begin{cases}
f(t) + k &\qquad \text{if taxpayer evades at time } t\\
f(t) + k(1-\tau) &\qquad \text{otherwise.}
\end{cases}
$$
Denote by $\pf(t)$ the profit made from evasion up to and including time $t$. Then
$$
\pf(t+1) = \begin{cases}
\pf(t) + \tau &\qquad \text{if taxpayer evades at time } t\\
\pf(t) &\qquad \text{otherwise.}
\end{cases}
$$
Denote by $\n(t)$ the number of times the taxpayer has evaded since the last audit. Then 
$$
\n(t+1) = \begin{cases}
\n(t) + 1 &\qquad \text{if taxpayer evades at time } t\\
\n(t) &\qquad \text{otherwise.}
\end{cases}
$$
Suppose the taxpayer is audited at time $t$ with probability $p$. Audits are independent of one another. If the taxpayer is audited then the taxpayer must pay $\lambda\tau \n(t+1)$. If $f(t+1) < \lambda\tau\n(t+1)$ then the taxpayer must pay the entire fortune $f(t+1)$ instead. Thus, if audited, $f$, $\pf$ and $\n$ are updated according to the rules
\begin{align*}
f(t+1) &\leftarrow f(t+1) - \min(f(t+1), \lambda\tau \n(t+1) )\\
\pf(t+1) &\leftarrow \pf(t+1) - \min(f(t+1), \lambda\tau \n(t+1))\\
\n(t+1) &\leftarrow 0
\end{align*}
It remains to give the condition for evasion. We assume that the taxpayer chooses to evade if $\pf(t) > 0$. In other words, if the profit made from evasion so far is positive, the taxpayer chooses to evade because evasion looks like a good bet. We also assume $f(0) = \n(0) = 0$ and $\pf(0) > 0$ (otherwise the taxpayer will never evade.)

\subsection{Difference Equations}\label{differenceequations}
The model can be described even more formally via the following system of equations
\begin{align*}
A_t &\sim \mathrm{Bernoulli}(p)\\
f(t+1) &= f(t) + k(1-\tau) + k\tau\delta_{\pf(t) > 0} - \min\left( f(t) + k(1-\tau) + k\tau\delta_{\pf(t) > 0},
\lambda\tau (\n(t) + \delta_{\pf(t) > 0}) \right)A_t\\
\pf(t+1) &= \pf(t) + \tau - \min\left( f(t) + k(1-\tau) + k\tau\delta_{\pf(t) > 0},
\lambda\tau (\n(t) + \delta_{\pf(t) > 0})\right)A_t\\
\n(t+1) & = (\n(t) + \delta_{\pf(t) > 0})(1-A_t)
\end{align*}
where $\delta_{\pf(t) > 0}$ is $1$ if $\pf(t) > 0$ and $0$ if $\pf(t) \le 0$, and $f(0) = \n(0) = 0$ and $\pf(0) > 0$. 

\section{Solution of the One-Taxpayer Model}\label{solution}
\subsection{}
In this section we describe the dynamics of the one-taxpayer model. If $\pf(t) \le 0$ at some time $t$, then the taxpayer never evades again, so we restrict to the case $\pf(t) > 0$. In this case, $\pf(t)$ cannot decrease until the taxpayer is audited, so between audits the taxpayer always evades.
\subsection{}
Suppose the taxpayer is audited at time $T_0$ and at time $T_1 > T_0$ and there are no audits in between. We consider the total profit made from evasion
$$\pf(T_1) - \pf(T_0)$$ 
between $T_0$ and $T_1$.
\subsection{}
In the first case, suppose $\lambda \tau \le k$. Then $f(t) \ge \lambda\tau\n(t)$ for all $t$. This follows by induction on $t$. It is true for $t=0$. If $f(t) \ge \lambda\tau\n(t)$ and $A_t=0$ then $f(t+1) = f(t) + k \ge \lambda\tau\n(t) + \lambda\tau = \lambda\tau\n(t+1)$. If $f(t) \ge \lambda\tau\n(t)$ and $A_t=1$ then $\n(t+1)=0$ and so $f(t+1) \ge \lambda\tau\n(t+1)$ in this case too. Therefore, in case $\lambda\tau \le k$, we have $f(t) \ge \lambda\tau\n(t)$ for all $t$ and
\begin{align*}
\pf(T_1) - \pf(T_0) &= \tau(T_1 - T_0) - \min(f(T_1) + k, \lambda\tau(\n(T_1) + 1))\\
&= \tau(T_1 - T_0) - \lambda\tau (\n(T_1) + 1)
\end{align*}
but $\n(T_1) + 1 = T_1 - T_0$ because the taxpayer evades at times $T_0 + 1, T_0 + 2, \ldots, T_1$ and so
$$
\pf(T_1) - \pf(T_0) = \tau(1-\lambda)(T_1 - T_0)
$$

\subsection{}\label{DpfDt}
In the second case, suppose $\lambda\tau > k$. Then $f(t) \le \lambda\tau \n(t)$ for all $t$. This follows by induction on $t$. It is true for $t=0$. If $f(t) \le \lambda\tau\n(t)$ and $A_t=0$ then $f(t+1) = f(t) + k \le \lambda\tau\n(t) + \lambda\tau = \lambda\tau\n(t+1)$. If $f(t) \le \lambda\tau\n(t)$ and $A_t=1$ then $f(t+1) = 0$ and so $0 = f(t+1) \le \lambda\tau\n(t+1)$ in this case too. Therefore, in case $\lambda\tau > k$ we have $f(t) \le \lambda\tau\n(t)$ for all $t$ and 
\begin{align*}
\pf(T_1) - \pf(T_0) &= \tau(T_1 - T_0) - \min(f(T_1) + k, \lambda\tau(\n(T_1) + 1))\\
&= \tau(T_1 - T_0) - (f(T_1) + k)
\end{align*}
but $f(T_1) + k = k(T_1 - T_0)$ because $f(T_0) = 0$ in this case (the taxpayer is audited at time $T_0$ by assumption and loses their entire fortune because $f(T_0-1) + k \le \lambda\tau(n(T_0-1)+1)$) and the taxpayer evades at times $T_0 + 1, T_0 + 2, \ldots, T_1$, and so
$$
\pf(T_1) - \pf(T_0) =(\tau -k)(T_1 - T_0)
$$
\subsection{}
Although $T_0$ and $T_1$ are random variables, we see that in all cases the quantity
$$
\frac{\pf(T_1)-\pf(T_0)}{T_1-T_0} = \tau - \min(k, \lambda\tau) = \begin{cases}\tau -k &\qquad \tau > k/\lambda\\ \tau(1-\lambda) &\qquad \tau \le k/\lambda \end{cases}
$$
is constant. This quantity represents the average change in $\pf$ between audits. Equivalently, it represents the average profit from evasion. We will use it extensively in this paper and denote it using the derivative-like symbol $\DpfDt$.

\subsection{}
We see that if $\tau > k$ then $\pf$ is an increasing function from one audit to the next, and therefore the taxpayer always evades. If $\tau < k$ then $\pf$ will eventually become negative and the taxpayer becomes compliant. In this case, we can calculate the expected amount of time which elapses until the taxpayer becomes compliant. Let $T_{\text{comp}}$ be the time at which $\pf$ first becomes negative. Each audit reduces $\pf$ by $\DpfDt \times \Delta T$ where $\Delta T$ is the time elapsed since the last audit. So $\pf$ will become negative as soon as there is an audit after time $-\pf(0)/\DpfDt$ has elapsed. The expected time until the next audit is a geometric random variable with a mean of $1/p$ and therefore
\begin{equation}\label{timetocomply}
E[T_{\text{comp}}] = -\pf(0)\left(\DpfDt\right)^{-1} + \frac{1}{p} = \begin{cases}
\frac{\pf(0)}{(k-\tau)} + \frac{1}{p} &\qquad \tau \ge k/\lambda\\
\frac{\pf(0)}{\tau(\lambda-1)} + \frac{1}{p} &\qquad \tau < k/\lambda
\end{cases}
\end{equation}
In particular, the expected time until the taxpayer becomes compliant is a linear function of $-\left(\DpfDt\right)^{-1}$.

\subsection{Examples}
The calculations of Section \ref{solution} can easily be tested by implementing the equations of Section \ref{differenceequations} on a computer.  Plots of $\pf(t)$ against $t$ from three example runs are shown in Figure \ref{fig1}. Here $k=0.4$ and $\lambda=1.5$. The first two graphs show results for $\tau = 0.02$ and $\tau = 0.39$. These taxpayers eventually become compliant but at different rates, as shown by the dotted lines with slope $\DpfDt$. The third taxpayer has $\tau = 0.42$. This tax rate is too high for the taxpayer ever to become compliant. 

\begin{figure}
\centering
\includegraphics[width = 6in]{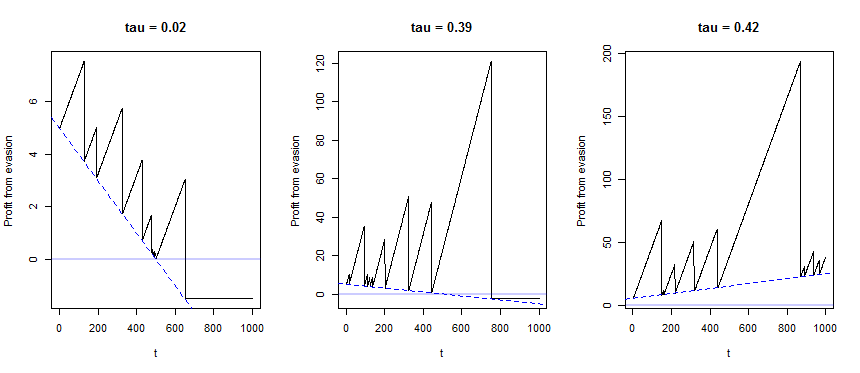}
\caption{Plots of $\pf(t)$ versus $t$ from simulations from the one-taxpayer model with $k=0.4$, $p=0.01$, $\pf(0)=5$, $\lambda=1.5$ and $\tau = 0.02, 0.39, 0.42$. The dotted lines are $y(t) = \pf(0) + \tau(1-\lambda)t$ (left) and $y(t) = \pf(0) + (\tau-k)t$ (middle and right).}\label{fig1} 
\end{figure}

\subsection{Measuring Evasion}
When $\tau < k$, the taxpayer always becomes compliant eventually. However, Equation \ref{timetocomply} shows that the expected time taken to reach compliance (which is the same as the number of times the taxpayer is expected to evade until becoming compliant) is a linear function of $-(\Delta \pf/\Delta T)^{-1}$ with positive coefficients. It is therefore reasonable to take $-(\Delta \pf/\Delta T)^{-1}$  as a measure of non-compliance.  It is likewise reasonable to take $\DpfDt$ as a measure of non-compliance because it measures how profitable evasion is for the taxpayer on average.

\subsection{}
The one-taxpayer model is interesting because it exhibits phenomena which can be seen in the real world but not in many models of tax evasion. 
A graph of $\DpfDt$ against $\tau$ for fixed values of $k$ and $\lambda$ is shown in Figure \ref{fig2}. 
Also sketched in Figure \ref{fig2} is a graph of $-(\Delta \pf/\Delta T)^{-1}$ against $\tau$. It will be useful to compare the U-shape of this graph with experimental results from 
Figures \ref{evaders_vs_tau_star} and \ref{evaders_vs_tau_ising}.

\begin{figure}
\centering
\includegraphics[width = 7in]{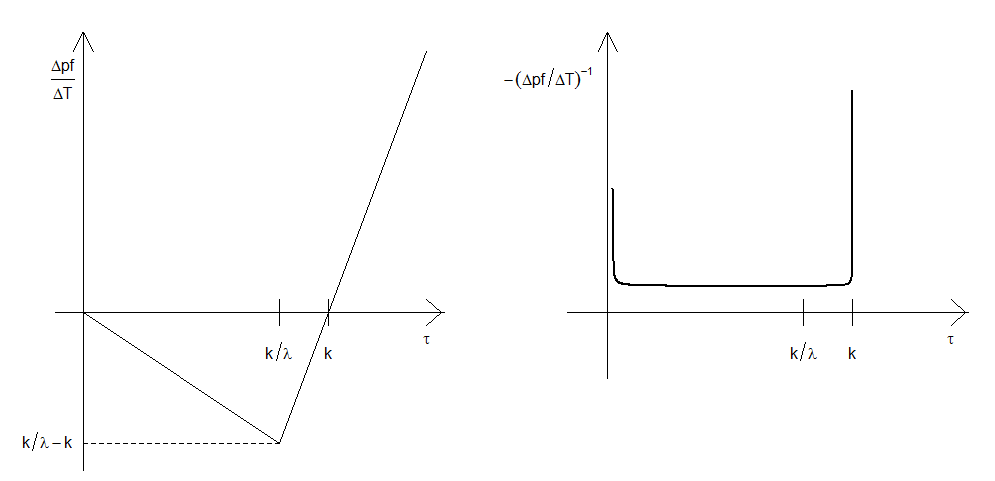}
\caption{Left panel: graph of $\DpfDt$ versus $\tau$. Evasion becomes profitable when $\tau > k$ and is minimised when $\tau = k/\lambda$. Right panel: graph of $-(\Delta \pf/\Delta T)^{-1}$. The shape of this graph should be compared with the experimental results in Figure \ref{evaders_vs_tau_star} and Figure \ref{evaders_vs_tau_ising}.}\label{fig2} 
\end{figure}

\subsection{Discussion}
If $\tau > k/\lambda$, increasing the tax rate causes compliance to decrease. This fact does not require any arguments about the morality of taxpayers. It is a consequence of the cost to the taxpayer of non-compliance being bounded below, which means that evasion becomes more profitable to the taxpayer when the tax rate is higher. 

\subsection{}
When $\tau < k$, the taxpayer becomes compliant. When $\tau < k/\lambda$, increasing the tax rate causes non-compliance to decrease. Non-compliance is minimised when $\tau = k/\lambda$. This optimal tax rate increases with the rate of saving and decreases when the penalties for evasion become harsher.

\subsection{Progressive taxation}
The result that a tax rate of $\tau = k/\lambda$ minimises non-compliance might help to justify progressive taxation (which is a form of taxation in which income over a certain threshold is taxed at a higher rate.) All other things being equal, it seems reasonable that a higher proportion of income will be saved if it is above what is needed for survival. This corresponds to a higher $k$ and therefore a higher $k/\lambda$. This argument is similar to a common argument in favour of progressive taxation based on marginal utility but has a different motivation; the model suggests that taxes should be progressive if the tax authority wishes to minimise non-compliance.

\section{A Network of Taxpayers}
\subsection{}
In the real world, there is more than one taxpayer. One of the original motives for creating the model of Section \ref{model} was to use it as a model for a single taxpayer operating within a network of taxpayers who share information. The idea of investigating such a model was inspired by the work of Zaklan, Westerhoff and Stauffer \cite{Zaklan} and the agent-based model of Korobow, Axtell and Johnson \cite{KAJ}. In this section, we construct a model for a network of taxpayers and present some experimental results.

\subsection{}
We assume that $N$ taxpayers are connected in an undirected network. Let $A_{xy}=1$ if $x$ and $y$ are connected and $A_{xy}=0$ otherwise. Assume that $A_{xx}=0$ for all $x$ (there are no self-loops.) Suppose the tax rate $\tau$, savings rate $k$ and probability of audit $p$ are fixed quantities in $(0,1)$, and there is a fixed penalty rate $\lambda > 1$. 

\subsection{}
For a taxpayer $x$, define the \emph{neighbourhood}
$$N(x) = \{x\} \cup \{y : A_{xy} = 1\}$$
to be the set of taxpayers consisting of $x$ and the neighbours of $x$. 

\subsection{}
Time proceeds in discrete steps $t = 0, 1, 2, \ldots$. Let $f(x,t)$ be the fortune of taxpayer $x$ at time $t$, $\pf(x,t)$ be the profit made by $x$ from evasion up to and including time $t$, and $\n(x,t)$ be the number of times $x$ has evaded since the last audit. Assume that taxpayers are audited independently of one another and that the probability that a given taxpayer $x$ is audited at a given time $t$ is $p$. Assume  $f(x,0)=\n(x,0)=0$ for all $x$ and that $f$, $\pf$ and $\n$ evolve for each taxpayer exactly as in the one-taxpayer model of Section \ref{model}, except for the criterion for choosing to evade.

\subsection{}
In the new model, the network structure enters through the rule which a taxpayer uses to decide whether to evade. A taxpayer is assumed to be aware of their own history and the histories of all of their neighbours and the decision to evade is based on their best guess at the expected profit from evasion given this information. Let $\n_{\mathrm{total}}(x, t)$ be the total number of times that $x$ evades between time $0$ and time $t$. A taxpayer $x$ chooses to evade if
\begin{equation}
\sum_{y \in N(x)}\pf(y,t) \Big/ \sum_{y \in N(x)}\n_{\mathrm{total}}(y,t) > 0.
\end{equation}
which is the same as
\begin{equation}\label{evasionrule1}
\sum_{y \in N(x)}\pf(y,t) > 0.
\end{equation}

\section{Analysis of the network-of-taxpayers model}
\subsection{}
The model for a network of taxpayers is no more difficult to implement than the one-taxpayer model. Simulations show that a network of taxpayers behaves in a similar way to a single taxpayer from a qualitative point of view. However, it seems to be much more difficult to calculate exactly what will happen for a network in the manner of Section \ref{solution}. The questions of interest are: Do the taxpayers eventually become compliant? How long does it take? Does the time until compliance vary from one taxpayer to another? 

\subsection{}
For a taxpayer $x$, $\pf(x,t)$ evolves in much the same way as in Section \ref{solution}. In particular, we need only consider the case $\DpfDt < 0$. But it is no longer guaranteed that $\pf(x,t)$ will become negative as $t \rightarrow \infty$; the taxpayer may become compliant before this happens. Nevertheless, if taxpayer $x$ is non-compliant, then $\pf(x,t)$ will decrease when $x$ is audited. Therefore, the taxpayers must eventually become compliant. 

\subsection{}
We make only a crude estimate of the time until compliance. Let $X$ be the set of taxpayers and $A \subset X$. (The case $A=N(x)$ is of particular interest.) Define 
$$\pf(A, t) = \sum_{x \in A}\pf(x,t).$$
Ignoring the randomness in the audits and assuming that every taxpayer evades at every time step, we have
$$\pf(x,t) \approx \pf(x,0) + \left(\DpfDt\right)t$$
and so
$$\pf(A,t) \approx \sum_{x \in A}\pf(x,0) + |A|\left(\DpfDt\right)t$$
which becomes negative at time
\begin{equation}\label{wrongtime}
T_{\mathrm{comp}} = \frac{\frac{1}{|A|}\sum_{x\in A}\pf(x,0)}{
-\left(\DpfDt\right)}.
\end{equation}
Therefore, if all the $\pf(x,0)$ are assumed to be equal, every subset of the taxpayers will become compliant at the same time. In particular, all the taxpayers are expected to become compliant at the same time. The time taken to become compliant does not depend on the network topology. 

\subsection{Example}\label{starnetwork}
The model was simulated for a  star-shaped network with ten nodes labelled with the integers $1$ to $10$. Node $1$ was connected to all other nodes and there were no other connections. Taking $\pf(x,0)=1$ for all $x$ and $k=0.4, \tau=0.3, \lambda=1.5, p=0.01$ the results of $50$ simulations are shown in Table \ref{starshaped}. For each node $x$, the time in the table is the number of the last iteration on which $x$ evaded.

\begin{table}
\centering
\begin{tabular}{|c|c|c|c|c|c|c|c|c|c|c|}
\hline
node &1 (centre)&2&3&4&5&6 &7 &8 &9 &10 \\
\hline
mean &420 &410 &394 &434 &435 &408 &415 &412 &408 &425 \\
\hline
sd &153 &239 &240 &248 &242 &237 &236& 254& 231& 237\\
\hline
\end{tabular}
\vspace{2mm}
\caption{Time until compliance for ten taxpayers in a star-shaped network for $50$ simulations of the model. See Section \ref{starnetwork} for details.}
\label{starshaped}
\end{table}

\subsection{}
There is no evidence of a difference between nodes in the times until the nodes become compliant. However, Equation \ref{wrongtime} gives an estimate of $1/(k-\tau) = 10$ for the time for a node to become compliant, which is clearly wrong. This is because Equation \ref{wrongtime} ignores the audit probability $p$, which influences the time until compliance. Giving an accurate estimate of the time until compliance seems to be a more difficult problem than in Section \ref{solution}, especially when the $\pf(x,0)$ are allowed to take arbitrary values.

\subsection{}
Figure \ref{evaders_vs_tau_star} shows the average number of evaders plotted against the tax rate for the star-shaped network with $10$ vertices and $k = 0.4, p = 0.01, \lambda=1.5$. The vertical line is at the value $\tau = k/\lambda$. Comparison with Figure \ref{fig2} shows that there is a wide range of parameters in the network case for which non-compliance is approximately minimised. All values of $\tau$ between about $0.1$ and $0.3$ seem to be equally good.

\begin{figure}
\centering
\includegraphics[width = 4in]{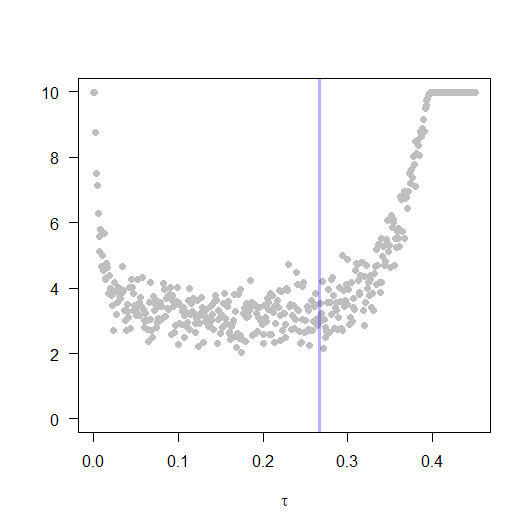}
\caption{Average number of evaders versus tax rate $\tau$ for a star-shaped network with $10$ taxpayers run for $1000$ iterations. The line is at $\tau = k/\lambda$, the tax rate which minimised non-compliance in the one-taxpayer model. Compare Figure \ref{fig2}.}\label{evaders_vs_tau_star} 
\end{figure}

\subsection{Example}\label{torus}
Zaklan, Westerhoff and Stauffer \cite{Zaklan} considered an Ising-like model in which taxpayers are connected in a square lattice. Korobow, Johnson and Axtell \cite{KAJ} considered a different model with a similar network topology but with diagonal connections. We simulated the model for $100$ taxpayers arranged in a square lattice. The lattice is wrapped around to form a torus, so for example the taxpayer at $(1,1)$ is adjacent to $(1,2)$, $(2,1)$, $(1,10)$ and $(10,1)$. This ensures that every taxpayer is adjacent to $4$ other taxpayers. As in the previous example, it is predicted that a tax rate of $\tau = k/\lambda$ will minimise non-compliance. 

\subsection{}
A plot of the average number of evaders against the tax rate $\tau$ is shown in Figure \ref{evaders_vs_tau_ising}. It can be seen that the shape is very similar to Figure \ref{evaders_vs_tau_star}, which is further evidence to suggest that the topology of the network has no influence on compliance behaviour in this model.

\begin{figure}
\centering
\includegraphics[width = 4in]{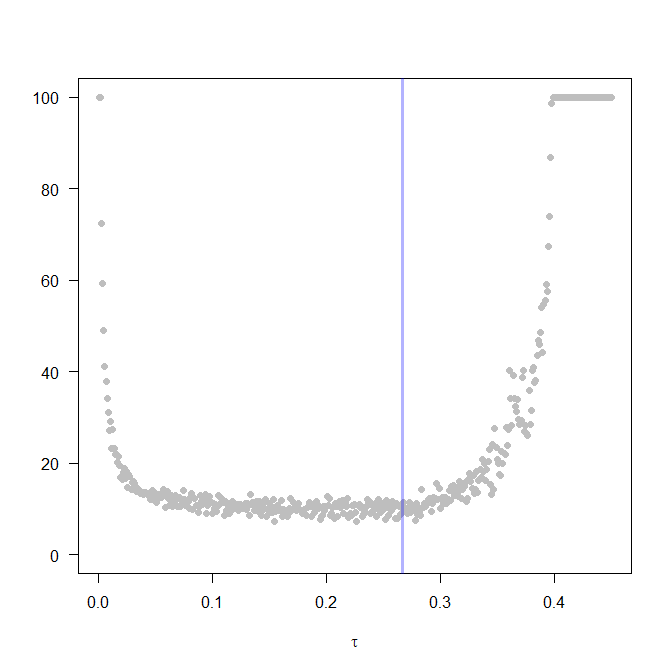}
\caption{Average number of evaders versus tax rate $\tau$ for a toroidal grid with $100$ taxpayers run for $1000$ iterations. The line is at $\tau = k/\lambda$, the tax rate which minimised non-compliance in the one-taxpayer model. Compare Figure \ref{fig2} and Figure \ref{evaders_vs_tau_star}.}\label{evaders_vs_tau_ising} 
\end{figure}

\subsection{Varying the probability of evasion}
We can make the taxpayer model even more like the Ising model by specifying a probability of evasion. The probability of taxpayer $x$ evading at time $t$ could be chosen to be an increasing function of $\sum_{y \in N(x)} \pf(x,t)$. By analogy with the Ising model, this could for example be taken to be
$$\exp\left(\beta\sum_{y \in N(x)} \pf(x,t) \middle/ \sum_{y \in N(x)} \n_{\mathrm{total}}(x,t)\right)$$
where $\beta > 0$ is a parameter which plays the role of the reciprocal of temperature. As $\beta \rightarrow \infty$, this reduces to (\ref{evasionrule1}). For smaller values of $\beta$, there is more randomness in the decisions of the taxpayers and as $t \rightarrow \infty$, the model approaches Bernoulli trials with a probability $\exp(\beta \DpfDt)$ of evasion. Otherwise, this choice of evasion probability does not affect the qualitative behaviour of the model when the parameters are held constant, so we choose not to discuss it further.

\subsection{Effect of heterogeneous $k$}
The model is still unrealistic in many ways. For example, real taxpayers do not inhabit a torus, and the savings rate $k$ undoubtedly varies over time and from person to person. There are many directions for further investigation. In this section we investigate what happens when each taxpayer is allowed to have an individual value for $k$. This follows the ideas of Korobow, Axtell and Johnson, who allowed various parameters to vary across taxpayers in their model.

\subsection{}
Under the conditions of Section \ref{torus}, suppose that the savings rates of the $100$ taxpayers are drawn independently from a $\mathrm{Beta}(2,3)$ distribution. This ensures that the mean savings rate is still about $0.4$. The tax rate is chosen to be $\tau = 0.3$ and the probability of audit is $p=0.1$. The model is run for $10000$ iterations. Everything else is kept the same as in Section \ref{torus}. We investigate the effect of varying $k$ on compliance. We expect that individuals with higher $k$ should be more compliant on average, but this might be tempered by network effects.

\subsection{}
It is of interest to see how many times a given taxpayer evades in the $10000$ iterations. Figure \ref{example_sociomatrix} shows a plot of a typical model run. The $100$ taxpayers are plotted in a $10\times 10$ grid and the darker squares indicate taxpayers who evade more often. 
The network is a torus, so the bottom row and top row are adjacent and the left column is adjacent to the right column. Note that the black and white regions in Figure \ref{example_sociomatrix} are almost contiguous, which is surprising because the savings rates of individual taxpayers were chosen at random. However, experiments with larger grids, for example $100 \times 100$, suggest that evaders tend to occur in small cross-shaped regions which are scattered at random throughout the network.

\begin{figure}[!htbp]
\centering
\includegraphics[width = 4in]{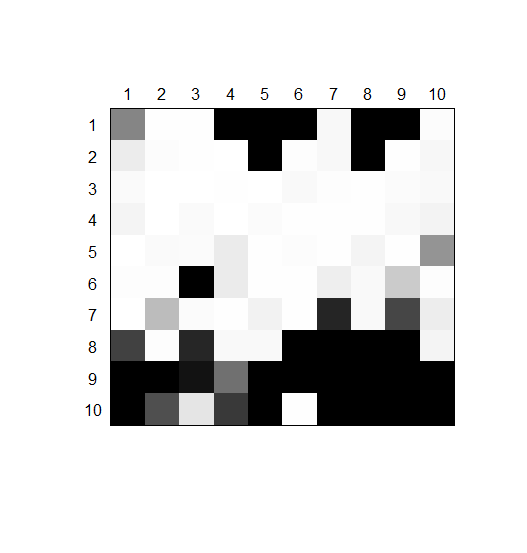}
\caption{Total number of evasions for each taxpayer in a $10 \times 10$ toroidal grid run for $10000$ iterations. The value of $k$ varies from taxpayer to taxpayer. The other parameters have the values $\lambda=1.5, \tau=0.3, p=0.1$.}\label{example_sociomatrix} 
\end{figure}

\begin{figure}[!htbp]
\centering
\includegraphics[width = 6in]{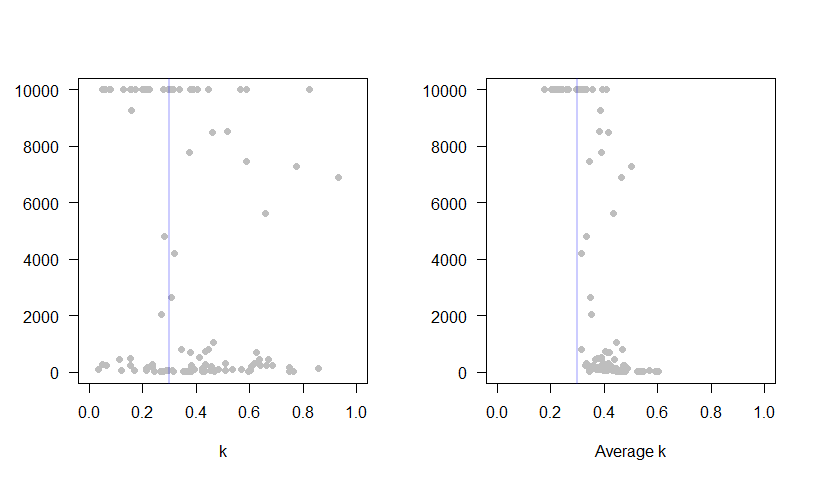}
\caption{Total number of evasions plotted against $k$ (left panel) and average $k$ in a five-taxpayer neighbourhood (right panel) for each taxpayer in a $10 \times 10$ toroidal grid. The value of $k$ varies from taxpayer to taxpayer. The other parameters have the values $\lambda=1.5, \tau=0.3, p=0.1$. The vertical lines are at $k=\tau$. }\label{savings_rate} 
\end{figure}

\subsection{}
The number of evasions for the $100$ taxpayers is plotted against the value of $k$ in the left panel of Figure \ref{savings_rate}. There seems to be no obvious pattern. In the right panel of Figure \ref{savings_rate}, the number of evasions is plotted against the average value of $k$, $k_{\mathrm{avg}} = \sum_{y \in N(x)} k(y)$ for the $100$ taxpayers $x$. It seems that taxpayers with $k_{\mathrm{avg}}(x) < \tau$ evade every time, but there is no obvious pattern for taxpayers with $k_{\mathrm{avg}}(x) > \tau$. Also, the majority of taxpayers either become highly compliant or highly evasive, with few in between the two extremes.

\section{Acknowledgements}
The author thanks D. Nagin and S. J. Spurr for providing a copy of their paper \cite{KNS}.

\section{Conclusion}
\subsection{}
We have given a new model for tax compliance. The model for a single taxpayer is simple enough to be exactly solvable and has two realistic properties: it predicts that sensible tax rates will eventually lead to compliance and that there is a realistic relationship between changes in the tax rate and changes in the level of non-compliance. We propose that variants of this single-taxpayer model might be useful as agents in an agent-based tax evasion model.

\bibliography{taxbib}

\begin{thebibliography}{1}

\bibitem{AS}
Michael~G. Allingham and Agnar Sandmo.
\newblock Income tax evasion: a theoretical analysis.
\newblock {\em Journal of Public Economics}, 1(3-4):323--338, 1972.

\bibitem{SP}
Mar\'{i}a~Jes\'{e}s Freire-Ser\'{e}n and Judith Panad\'{e}s.
\newblock {D}o {H}igher {T}ax {R}ates {E}ncourage/{D}iscourage {T}ax
  {C}ompliance?
\newblock {\em Modern Economy}, 4(12):809--817, 2013.

\bibitem{KAJ}
A.~Korobow, C.~Johnson, and R.~L. Axtell.
\newblock {A}n {A}gent-based {M}odel of {T}ax {C}ompliance with {S}ocial
  {N}etworks.
\newblock {\em National Tax Journal}, 60(3):589--610, 2007.

\bibitem{KNS}
D.~Nagin S.~Klepper and S.~Spurr.
\newblock {T}ax {R}ates, {T}ax {C}ompliance, and the {R}eporting of
  {L}ong-{T}erm {C}apital {G}ains.
\newblock {\em Public Finance/Finances Publiques}, 46(2):236--251, 1991.

\bibitem{Seibold}
G\"{o}tz Seibold and Michael Pickhardt.
\newblock {L}apse of time effects on tax evasion in an agent-based econophysics
  model.
\newblock {\em Physica A: Statistical Mechanics and its Applications},
  392(9):2079--2087, 2013.

\bibitem{Yitzhaki}
Shlomo Yitzhaki.
\newblock Income tax evasion: a theoretical analysis.
\newblock {\em Journal of Public Economics}, 3(2):201--202, 1974.

\bibitem{Zaklan}
Georg Zaklan, Frank Westerhoff, and Dietrich Stauffer.
\newblock {A}nalysing tax evasion dynamics via the {I}sing model.
\newblock {\em Journal of Economic Interaction and Coordination}, 4(1):1--14,
  2009.

\end{thebibliography}
\bibliographystyle{plain}

%
%
%

\end{flushleft}

\end{document}